\documentclass[journal]{IEEEtran}

\usepackage{graphicx}
\usepackage[tbtags]{amsmath}
\usepackage{amssymb}
\usepackage{array}
\usepackage{bm}
\usepackage{color}
\usepackage[caption=false]{subfig}
\usepackage[hidelinks]{hyperref}

\newcommand\ddfrac[2]{\frac{\displaystyle #1}{\displaystyle #2}}
\DeclareMathOperator*{\argmin}{arg\,min}
\DeclareMathOperator*{\argmax}{arg\,max}

\begin{document}

\title{Probability-Based Optimal Control Design for \\ Soft Landing of Short-Stroke Actuators}

\author{%
	Eduardo~Moya-Lasheras,~\IEEEmembership{Student Member,~IEEE,}
	Edgar~Ramirez-Laboreo,~\IEEEmembership{Student Member,~IEEE,}
	and~Carlos~Sagues,~\IEEEmembership{Senior Member,~IEEE}%
	\thanks{Manuscript received April 8, 2019; accepted May 10, 2019. Manuscript received in final form May 20, 2019. This work was supported in part by the Ministerio de Ciencia Innovaci\'on y Universidades, Gobierno de Espa\~na - European Union under project RTC-2017-5965-6 of subprogram Retos-Colaboraci\'on, in part by the Ministerio de Educaci\'on, Cultura y Deporte, Gobierno de Espa\~na under grant FPU14/04171, in part by DGA Scholarship Orden IIU/1/2017 confined within Programa Operativo FSE Arag\'on 2014-2020, and in part by project DGA-T45\_17R/FSE.}
	\thanks{The authors are with the Departamento de Informatica e Ingenieria de Sistemas (DIIS) and the Instituto de Investigacion en Ingenieria de Aragon (I3A), Universidad de Zaragoza, 50018 Zaragoza, Spain (email: emoya@unizar.es; ramirlab@unizar.es; csagues@unizar.es).}
	\thanks{\textcolor{red}{This is the accepted version of the manuscript: E. Moya-Lasheras, E. Ramirez-Laboreo and C. Sagues, ``Probability-Based Optimal Control Design for Soft Landing of Short-Stroke Actuators," in IEEE Transactions on Control Systems Technology, vol. 28, no. 5, pp. 1956--1963, Sept. 2020, doi: 10.1109/TCST.2019.2918479. \textbf{Please cite the publisher's version}. For the publisher's version and full citation details see:\\
	\protect\url{https://doi.org/10.1109/TCST.2019.2918479}. 
	}}
	\thanks{© 2019 IEEE.  Personal use of this material is permitted.  Permission from IEEE must be obtained for all other uses, in any current or future media, including reprinting/republishing this material for advertising or promotional purposes, creating new collective works, for resale or redistribution to servers or lists, or reuse of any copyrighted component of this work in other works.}
}

\maketitle

\makeatletter

\begin{abstract}
	The impact forces during switching operations of short-stroke actuators may cause bouncing, audible noise and mechanical wear. The application of soft-landing control strategies to these devices aims at minimizing the impact velocities of their moving components to ultimately improve their lifetime and performance. In this paper, a novel approach for soft-landing trajectory planning, including probability functions, is proposed for optimal control of the actuators. The main contribution of the proposal is that it considers uncertainty in the contact position and hence the obtained trajectories are more robust against system uncertainties. The problem is formulated as an optimal control problem and transformed into a two-point boundary value problem for its numerical resolution. Simulated and experimental tests have been performed using a dynamic model and a commercial short-stroke solenoid valve. The results show a significant improvement in the expected velocities and accelerations at contact with respect to past solutions in which the contact position is assumed to be perfectly known.
\end{abstract}

\section{Introduction}
\IEEEPARstart{E}lectromechanical actuators are devices with movable parts operated by a source of electrical energy. In particular, switch-type actuators are characterized for having a limited range of motion, e.g. reluctance actuators (solenoid valves, relays, contactors), or microelectromechanical system (MEMS) switches. For many of these devices, the impact forces during switching operations cause bounces, audible noise, or mechanical wear \cite{Montanari2003,Peterson2004}. The design of soft-landing strategies is of great interest for these devices, as it permits enhancing their service life, broadening their scope of applications, or replacing more complex and expensive actuators.

Regarding the control of MEMS switches, the feedback design is problematic in many cases, because sensor data is noisy or unavailable, and dynamics is very fast \cite{Borovic2005}. Therefore, some works focus on designing waveforms that aim at reducing the impact velocities \cite{Sumali2007} in an open-loop fashion. To compensate for fabrication variability, \cite{Allen2008a} proposes to refine the design of the waveform via reliability-based design optimization and Monte Carlo simulations, which require great computational effort. Alternatively, \cite{Blecke2009} proposes a simple learning control to iteratively reduce contact bouncing.

For specific reluctance actuators, several works tackle the soft-landing problem. One approach is to design feedback control strategies to track a predefined position trajectory \cite{Mercorelli2016,Eyabi2006,Chladny2008,Mercorelli2012}. In the case that the position cannot be measured in real time, some works focus on the design of estimators of the state  \cite{Moya-Lasheras2017,Braun2018} or other time-dependent variables of interest \cite{Ramirez-Laboreo2019}. To improve the robustness, some of them also propose cycle-to-cycle learning-type controllers to adjust the feedback controller \cite{Peterson2004,Benosman2015} or the feedforward signal \cite{Hoffmann2003,Tsai2012,Ramirez-Laboreo2017a}. However, for certain simple low-cost reluctance actuators, the soft-landing problem is not satisfactorily solved. On the one hand, their dynamics are fast, and highly nonlinear, which makes the design of position observers a challenging task. On the other hand, their position cannot be measured in real time, either because a sensor cannot be placed or because it is much more expensive than the device.

The design of a tracking trajectory and its corresponding input signal is a key point for both feedback and feedforward control. The generation of trajectories is discussed in previous works for different actuators, and it is common to assume that errors in models, observers and measurements are negligible. On that assumption, soft landing is achieved by setting as bound conditions the final velocity and acceleration equal to zero. Trajectory planning is therefore focused on finding feasible solutions \cite{Koch2004} or on optimizing some particular variables, e.g. transition time \cite{Gluck2011} or mean power consumption \cite{Fabbrini2012}. However, in practice, and specially for low-cost actuators, the system representation is not perfect and therefore the generated optimal input signals do not result in real soft landing when the control is implemented.

In this brief paper, a novel approach for soft-landing open-loop control is developed. The contribution of our proposal is the addition of probability functions in the optimal control problem for trajectory planning. Specifically, uncertainty in the contact position is included and the soft-landing optimal control is formulated in order to minimize the expectations of the contact velocity and acceleration. Furthermore, the advantages of utilizing the electrical current as the control input for reluctance actuators are discussed and, in consequence, the optimization of the current signal is included in the formulation of the problem. Simulated and experimental tests have been carried out to analyze the applicability of the designed trajectories in an open-loop controller and the improvement due to the inclusion of uncertainty.

\section{Problem statement}\label{Generalized system dynamics}

The first step of the proposed method is the definition of the motion dynamics of a generic device. The proposed representation is a generalized lumped parameter model, accounting for the mechanical subsystem and any other dynamics that influences it, e.g. electrical or magnetic. It can be expressed as a set of two or more differential equations,
\begin{subequations}\label{state_space_generic}
	\begin{align}
		\hspace{-3mm} \dot z(t)           & \!=\! v(t), \label{x1x2}                                                                                                                     \\
		\hspace{-3mm} \dot v(t)           & \!=\! f_v\big(z(t), v(t), \bm \alpha(t) \big) + G_v\big(z(t), v(t), \bm \alpha(t)\big) \, \bm u(t), \label{eq: dot v}                        \\
		\hspace{-3mm} \bm{\dot \alpha}(t) & \!=\! \bm{f_\alpha}\big(z(t), v(t), \bm \alpha(t)\big) + \bm{G_\alpha}\big(z(t), v(t), \bm \alpha(t)\big) \, \bm u(t), \label{eq: dot alpha}
	\end{align}
\end{subequations}
where $z$ and $v$ are the position and velocity of the movable part. Additional state variables are condensed in the vector $\bm \alpha \in \mathbb R^{n-2}$, being $n \geq 2$ the order of the dynamical system. The input vector $\bm u$ may affect directly the acceleration $\dot v$ or the dynamics of $\bm \alpha$, which in turn influences $\dot v$. The model is general enough to include a wide range of actuators. This set of equations can be compactly expressed as
\begin{equation}\label{system dynamics}
	\dot{\bm x}(t) = \bm f\big(\bm x(t) \big) + \bm G\big(\bm x(t)\big) \, \bm u(t),
\end{equation}
where ${\bm x = } \begin{bmatrix} {z} & {v} & {\bm \alpha^\mathsf T }\end{bmatrix}{^\mathsf T}$. Secondly, the soft-landing trajectory planning is formulated as a standard optimal control problem, where the cost is a functional of a scalar function $V$ of the state and the input, i.e.
\begin{equation}
	J = \int_0^{t_\mathrm f} V\big(\bm x(t), {\bm u}(t)\big) \, \mathrm dt,
\end{equation}
where $t_\mathrm f$ is the final time. The definition of $V$ for soft landing is specified in the following section. The optimization problem is then solved via Pontryagin's Minimum Principle \cite{Naidu2003}. Provided that $\bm \lambda$ is the costate vector, and that the Hamiltonian is\looseness=-1
\begin{equation}\label{eq: Ham}
	\hspace{-1mm}\mathcal H\big(\bm x(t), \bm \lambda(t) , \bm u(t)\big)  = V\big(\bm x(t),u(t)\big) + \bm \lambda^\mathsf T \bm f\big(\bm x(t), \bm u(t)\big),
\end{equation}
the optimal control $\bm u^*$ must satisfy the following condition,
\begin{align}\label{eq: minH}
	\bm u^* & (t) =\bm u^*\big(\bm x^*(t), \bm \lambda^*(t)\big) \quad \mathrm{s.t.} \nonumber                                                                                 \\
	        & \mathcal H\big(\bm x^*(t), \bm \lambda^*(t) , \bm u^*\big) \leq \mathcal H\big(\bm x^*(t), \bm \lambda^*(t) , \bm u\big), \, \forall \bm u \in \bm{\mathcal{U}},
\end{align}
where $\bm x^*$ and $\bm \lambda^*$ are the optimal state and costate vectors, and $\bm{\mathcal{U}}$ is the set of permissible values for $\bm u$. Then, the optimal control problem is reformulated as a two-point boundary value problem (BVP), with the differential equations for $\bm x^*$ and $\bm \lambda^*$,
\begin{subequations}
	\begin{align}\label{state-costate}
		\dot{\bm{x}}^*       & = \bm f\big(\bm x^*(t), \bm u^*(\bm x^*(t), \bm \lambda^*(t))\big),                                          \\
		\dot{\bm{\lambda}}^* & = -\partial \mathcal H\big(\bm x^*, \bm \lambda^*, \bm u^*(\bm x^*, \bm \lambda^*) \big) / \partial \bm x^*,
	\end{align}
\end{subequations}
subject to a set of $2 \, n$ boundary conditions. The boundary conditions for $t=0$ correspond to the beginning of the motion from initial position $z_0$,
\begin{align}
	z(0)= z_0, &  & \frac{\mathrm d z}{\mathrm dt}(0)=v(0)= 0, &  & \ldots, &  & \frac{\mathrm d^{n-1} z}{\mathrm dt^{n-1}}(0) = 0.
\end{align}

Given the assumption that the model is a perfect representation of the dynamical system, soft landing could be achieved by setting to zero the final velocity $v(t_\mathrm f)$, and higher position derivatives if $n>2$, as boundary conditions. However, the model is always a simplification of the system. To account for expected uncertainty and obtain a more conservative trajectory, the actual contact position is assumed a random variable $Z_\mathrm{c}$. Since the contact position is random, so it is the contact instant $T_\mathrm c$, velocity $V_\mathrm c$ and other state variables, and they cannot be set as boundary conditions. Therefore, the boundary conditions for $t=t_\mathrm f$ correspond to a free-final state, except for the final position, which is set to $z_\mathrm f$,
\begin{equation}\label{boundary}
	\begin{aligned}
		z(t_\mathrm f)= z_\mathrm f, &  & \lambda_2(t_\mathrm f) = 0, &  & \ldots, &  & \lambda_n(t_\mathrm f) =  0.
	\end{aligned}
\end{equation}

As the actual contact position is unknown, the solution does not terminate when $z = Z_\mathrm c$. Instead, the choice of $z_\mathrm f$ establishes the probability that the contact occurs before $t_\mathrm f$. E.g., in the case that the contact position is a normal deviate ($Z_\mathrm c \sim \mathcal N(\mu_z, {\sigma_z}^2)$), the final position could be set as its expectation $\mu_z$, which means that there would be a $50 \ \%$ probability of $T_\mathrm c \leq t_\mathrm f$. Alternatively, setting $z_\mathrm f = \mu_z + 3\, \mathrm{sgn}(V_\mathrm c) \, \sigma_z$ would guarantee a contact with a $99.87 \ \%$ confidence. This is preferable because, by definition, the trajectories beyond $t_\mathrm f$ are not optimized. Note also that the differential equation \eqref{system dynamics} represents the unconstrained system dynamics assuming the contact has not happened yet ($t < T_\mathrm c$). As the contact velocity is not affected by the dynamics after that event, there is no need to model the dynamical system for the case $t > T_\mathrm c$.

\section{Soft-landing cost functional}\label{Cost functional}
In this section, a cost $J$ is defined to obtain an optimal position trajectory and its correspondent input signal, given the assumption that the contact position is random. The total cost functional is divided into several terms $J_i$,
\begin{equation}
	J  = \sum_i J_i = \sum_{i} \int_0^{t_\mathrm f} V_i\big(\bm{x}(t), \bm u(t)\big) \, \mathrm dt,
\end{equation}
and the functions $V_i$ are defined in the following subsections.

In many devices there are two asymmetrical switching operations, depending on the direction of movement. The optimal control problem is formulated to be solved separately for each type of operation. Nevertheless, the following reasonings and expressions are generalized in order to be used for both.

\subsection{Expected contact velocity}\label{expected contact velocity}
In an elastic collision, the bouncing velocity depends on the velocity just before contact. Therefore, in order to reduce both impact forces and bouncing, the expected contact velocity should be minimized. The actual contact position is assumed a random variable $Z_\mathrm c$ with a probability density function (PDF) $f_{Z_\mathrm c}(z)$. The PDF can be of any form, provided that it is differentiable on the time interval,
\begin{equation}
	\exists \, \partial^2 f_{Z_\mathrm c}(z) / \partial {z}^2, \ \forall \, {z}(t) \ \mathrm{s.t.} \ t \, \in [0, \ t_\mathrm f].
\end{equation}

Since the contact position is a random variable, the contact instant $T_\mathrm c$ is also random,
\begin{equation}
	T_\mathrm c = \big( t \mid z(t) = Z_\mathrm c \big).
\end{equation}

Its probability density function $f_{T_\mathrm c}(t)$ can be derived from $f_{Z_\mathrm c}({z})$. As an intermediate step, let us define $P_\mathrm c$ as the probability that at an arbitrary instant $t\geq0$ the contact has already occurred, i.e.
\begin{equation}\label{Pc}
	P_\mathrm c (t) = \left| \int_{z(0)}^{z(t)} \! f_{Z_\mathrm c}(z) \, \mathrm dz \right|.
\end{equation}

Depending on the motion direction, ${z}(t) \leq z(0)$ or ${z}(t) \geq {z}(0)$. The absolute value simply ensures that the probability is non-negative for both cases. Moreover, integration by substitution permits transforming \eqref{Pc} into a time integral,
\begin{equation}\label{Pc2}
	P_\mathrm c (t) =  \int_{0}^{t} \! \left| f_{Z_\mathrm c}\big(z(\tau)\big) \, \dot{z}(\tau) \right| \, \mathrm d\tau.
\end{equation}

The absolute value is moved inside the integral, which is only possible under the assumption that the integrand is always non-negative or non-positive. As the PDF $f_{Z_\mathrm c}(z)$ is non-negative by definition, the position ${z}$ must be a monotonic function of time. This condition might seem restrictive, but it is completely reasonable. If it is not satisfied, there would be at least one time interval in which the movable part goes backwards, away from the final position. This is clearly not an expected behavior in an optimal trajectory.

From \eqref{Pc2} and \eqref{x1x2}, the PDF $f_{T_\mathrm c}(t)$ can be calculated as
\begin{equation}\label{fTc}
	f_{T_\mathrm c}(t) = \mathrm dP_\mathrm c(t) / \mathrm dt = \left|v(t)\right| \, f_{Z_\mathrm c}\big(z(t)\big).
\end{equation}

The contact velocity is a function of a random variable, $V_\mathrm{c} = v(T_\mathrm c)$, and thus it is unknown. However, its expected value can be expressed as a conditional expectation,
\begin{equation}\label{exp_vel_2}
	\mathrm{E}[V_\mathrm{c}] = \mathrm{E}[ \, {v}(T_\mathrm c) \mid 0 \leq T_\mathrm c \leq t_\mathrm f \, ],
\end{equation}
which can be calculated as
\begin{equation}\label{EVc_2}
	\mathrm{E}[V_\mathrm{c}] = \frac {\int_{0}^{t_\mathrm f} \, {v}(t) \, f_{T_\mathrm c}(t) \, \mathrm dt} {P(0 \leq T_\mathrm c \leq t_\mathrm f)},
\end{equation}
and, substituting \eqref{fTc} into \eqref{EVc_2}, the final expression is
\begin{equation}\label{EVc_F}
	\mathrm{E}[V_\mathrm{c}] = \int_{0}^{t_\mathrm f} \! \frac {1}{P(0 \leq T_\mathrm c \leq t_\mathrm f)} \, {v}(t)\, \left|v(t)\right| \, f_{Z_\mathrm c}\big(z(t)\big) \, \mathrm dt.
\end{equation}

Note that $P(\,\cdots)$ is a probability constant that does not depend on the state trajectory. Therefore, the previous expression has the form of a standard optimal control performance index. Since the goal is to minimize the absolute value of the contact velocity, the proposed cost functional $J_1$ is proportional to the expectation of $\left|V_\mathrm{c}\right|$, and as the velocity cannot change sign, it can be expressed as
\begin{equation}\label{J_1_initial}
	J_1 =  w_1 \, \left| \mathrm{E}\left[V_\mathrm{c}\right] \right| = w_1 \, \mathrm{sgn}(V_\mathrm{c}) \, \mathrm E[V_\mathrm c],
\end{equation}
where $w_1$ is a weight constant. This constant and the following ones determine the importance of each cost functional and should be chosen accordingly. Substituting \eqref{EVc_F} into \eqref{J_1_initial} the final expression is obtained,
\begin{equation}\label{J_1}
	\begin{aligned}
		J_1 & = \int_0^{t_\mathrm f} V_1(\bm x(t)) \, \mathrm dt, & V_1(\bm x(t)) & =  \frac{w_1 \, {v}^2(t) \, f_{Z_\mathrm c}({z}(t))} {P(0 \leq T_\mathrm c \leq t_\mathrm f)}.
	\end{aligned}
\end{equation}

Note that the absolute values are removed because $\mathrm{sgn}\left(v\right) = \mathrm{sgn}\left(V_c\right)$ except when $v = 0$, in which case $V_1 = 0$.

\subsection{Expected bounced acceleration}\label{expected acceleration}
In the case that the system is second order ($n = 2$), the acceleration can be directly controlled by $\bm u$, and therefore it is sufficient to minimize the expected contact velocity. However, in most cases, $n>2$ and position derivatives of higher order (acceleration, jerk, jounce...) should be minimized as well to achieve soft landing. The reason is that, if their sign is opposite to the motion direction, they tend to separate the movable part from the final position, even in a completely inelastic collision. In this subsection, the cost functional for the minimization of the bounced acceleration is derived. In the rare cases in which $n>3$, the same line of reasoning should be followed for higher order derivatives of the position.

As stated in \eqref{eq: dot v}, the acceleration ${\dot v}$ {may depend} on the velocity, which can change abruptly in the contact instant. Thus, the bounced acceleration $a_\mathrm b$ after contact should be calculated from the bounced velocity $v_\mathrm b$,
\begin{multline}
	a_\mathrm b\big(z(t), v_\mathrm b(t), \bm \alpha(t), \bm u(t)\big) = f_v\big(z(t),v_\mathrm b(t), \bm \alpha(t)\big) \\ + G_v\big(z(t),v_\mathrm b(t), \bm \alpha(t)\big) \, \bm u(t).
\end{multline}

It is assumed that the delay between the impact and the bouncing is negligible, i.e., the elasto-plastic dynamics of the contact are much faster than the dynamics of the armature during unconstrained motion. Then, if contact occurs at $t$, the bounced velocity, in the most general form, is a function of the state and the input at that instant. The function depends on the loss of kinetic energy at impact, so the boundaries are $v_\mathrm b=-v$ (no energy loss) and $v_\mathrm b = 0$ (complete energy loss). In the case there is no accurate model of the bouncing phenomenon, it is possible to conservatively estimate $v_\mathrm b$ as
\begin{align}
	\hat v_\mathrm b\big(\bm x(t), \bm u(t)\big) = & \, \argmax_{v_\mathrm b} \, -\mathrm{sgn}(V_\mathrm c) \, a_\mathrm b\big(z(t), v_\mathrm b, \bm \alpha(t), \bm u(t)\big) \nonumber \\
	                                               & {\ \mathrm{s.t.}} {\;  v_\mathrm b \,v(t) \leq 0, \ \left| v_\mathrm b \right| \leq \left| v(t) \right|.}
\end{align}

It is important to notice that the bounced acceleration is only detrimental in the direction that separates the armature from the final position, i.e. in the opposite direction of the velocity. Taking that into account, the saturated bounced acceleration $a_\mathrm{b,sat}$ is defined as an auxiliary variable, in the case of contact,
\begin{equation}\label{a_sat}
	a_\mathrm{b,sat}\big(\bm x(t), \bm u(t)\big) = \left\lbrace
	\begin{array}{rrl}
		a_\mathrm b, & a_\mathrm b \, V_\mathrm c \leq 0, & \text{(take off)} \\
		0,           & a_\mathrm b \, V_\mathrm c > 0.    & \text{(hold)}     \\
	\end{array}\right.
\end{equation}

Note that the saturation is required for calculating the cost functional, the unconstrained acceleration $\dot v$ is still calculated from \eqref{eq: dot v}. Furthermore, to numerically solve the problem, $a_\mathrm{b,sat}$ is recommended to be differentiable, condition not met in \eqref{a_sat}. A smooth saturation function should be used instead.

The bounced acceleration at contact depends on the contact instant, which was defined in the previous subsection,
\begin{equation}
	{A_\mathrm c =  a_\mathrm{b,sat}\big(\bm x(T_\mathrm c), \bm u(T_\mathrm c)\big).}
\end{equation}
As was the case for the contact velocity, $\mathrm{E}[A_\mathrm{c}]$ is defined as a conditional expectation,
\begin{equation}
	\mathrm{E}[A_\mathrm{c}] = \mathrm{E}[ \, {a_\mathrm{b,sat}\big(\bm x(T_\mathrm c), \bm u(T_\mathrm c)\big)} \ | \ 0 \leq T_\mathrm c \leq t_\mathrm f \, ],
\end{equation}
which {can be computed as}
\begin{equation}\label{EAc_F}
	\mathrm{E}[A_\mathrm{c}] = \int_{0}^{t_\mathrm f} \frac {\left|v(t)\right| \, a_\mathrm{b,sat}\big(\bm x(t), \bm u(t)\big) \, f_{Z_\mathrm c}\big(z(t)\big)}{P(0 \leq T_\mathrm c \leq t_\mathrm f)} \, \mathrm dt.
\end{equation}

In this case, the objective is to minimize the absolute value of $A_\mathrm c$. In consequence, the cost functional term should be
\begin{equation}\label{J_2_initial}
	J_2 = w_2 \, \left|\mathrm E[A_\mathrm c]\right| = - w_2 \, \mathrm{sgn}(V_\mathrm{c}) \, \mathrm E[A_\mathrm c],
\end{equation}
where $w_2$ is another weight term. Substituting \eqref{EAc_F} into \eqref{J_2_initial}, the final expression is obtained,
\begin{equation}\label{J_2}
	\begin{gathered}
		J_2 = \int_0^{t_\mathrm f} V_2(\bm x(t),  {\bm u(t)}) \, \mathrm dt,\\
		\hspace{-1.5mm}V_2(\bm x(t),  {\bm u(t)}) \!=\! - \frac{w_2 \, v(t) \, {a_\mathrm{b,sat}\big(\bm x(t), \bm u(t)\big)} \, f_{Z_\mathrm c}\big(z(t)\big)}{P(0 \leq T_\mathrm c \leq t_\mathrm f)}.
	\end{gathered}
\end{equation}

\subsection{Regularization term}\label{di}
Up until this point, the Hamiltonian {\eqref{eq: Ham}} is linear in $\bm u$ because $V_1$ does not depend on $\bm u$, and $V_2$ is linear in $\bm u$---except when saturating {\eqref{a_sat}}. Thus, the optimal control \eqref{eq: minH} has discontinuities, which means the problem is ill-defined, complicating its numerical resolution. To circumvent this, it is common to add as a regularization term a quadratic expression with respect to $u$, making the optimal control continuous. We propose\looseness=-1
\begin{align}\label{J_3}
	J_3 = \int_0^{t_\mathrm f} V_3(t) \, \mathrm dt, &  & V_3= w_3 \, \left(\frac{\mathrm d h\big(\bm x(t)\big)}{\mathrm dt}\right)^2,
\end{align}
where $w_3$ is an additional weight term and $h$ is an arbitrary signal whose quadratic derivative is minimized. The cost functional $V_3$ can be expressed as
\begin{equation}\label{V_3}
	\hspace{-1mm} V_3(\bm x(t), \bm u(t)) \!=\! w_3 \left(\frac{\mathrm \partial h(\bm x)}{\partial \bm x} \Big( \bm f\big(\bm x(t)\big) + \bm G\big(\bm x(t)\big) \, \bm u(t) \Big)\right)^2,
\end{equation}
which is quadratic in $\bm u$, and thus serves as a regularization term to avoid solving an ill-defined optimal control.

In particular, for reluctance actuators, $ h(\bm x)$ can be the current through the coil, as reducing its derivative is advantageous in the case the system is going to be controlled by using the optimal current signal as reference or input (the motivation for this decision is discussed in the following section). The current signal obtained this way is less steep and therefore easier to follow accurately in the implementation. Therefore, $V_3$ is not only useful for the numerical resolution of the problem, because reducing the current derivative is also desirable from a purely theoretical standpoint. However, it is not possible to minimize simultaneously the current derivative and the expected contact velocity and acceleration, so there is a trade-off which depends on the chosen cost weights.
\section{Reluctance actuator}\label{sec: Reluctance actuator}
To analyze the optimal control proposal, a real reluctance actuator and its dynamic model is used for simulations and experiments. To improve the readability, from this point forward the time dependence of the variables is omitted.

\subsection{Description of actuator and dynamic model}\label{subsec: model}
\begin{figure}
	\includegraphics[height=3.5cm]{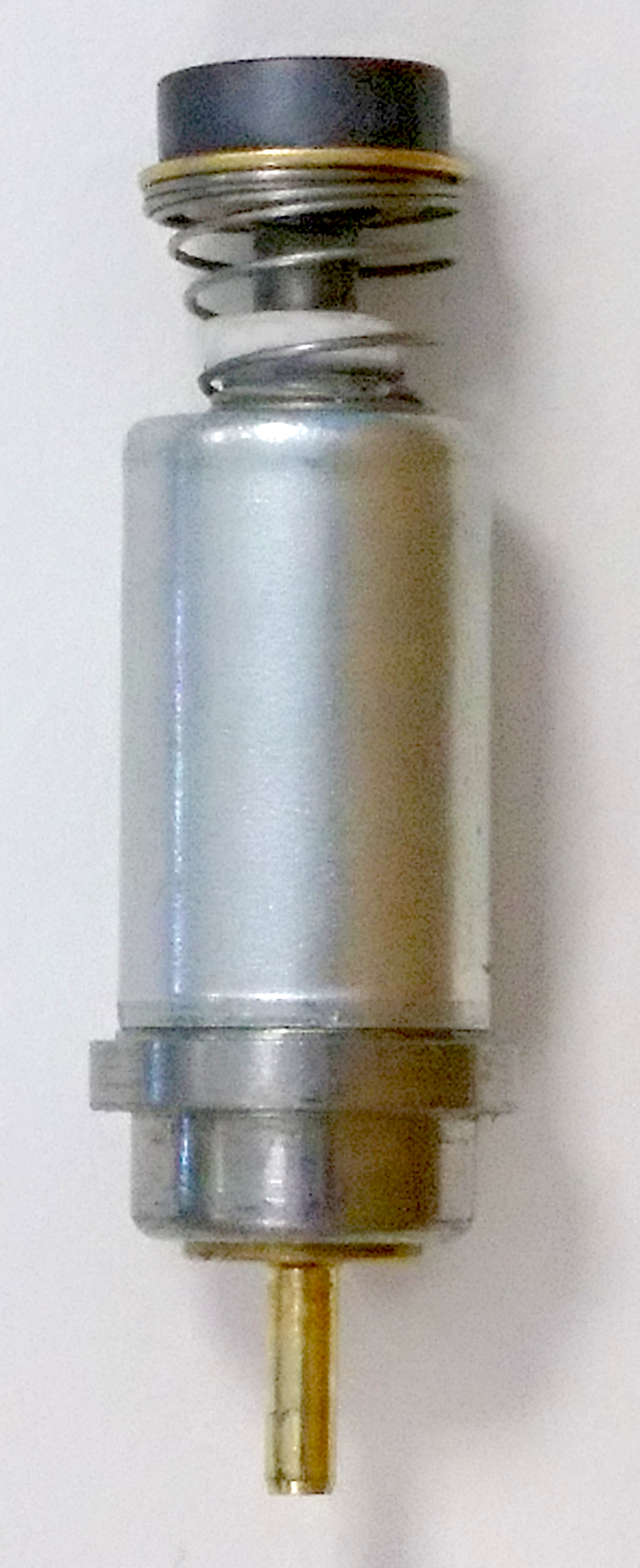} \hspace{\fill}
	\includegraphics[height=3.5cm]{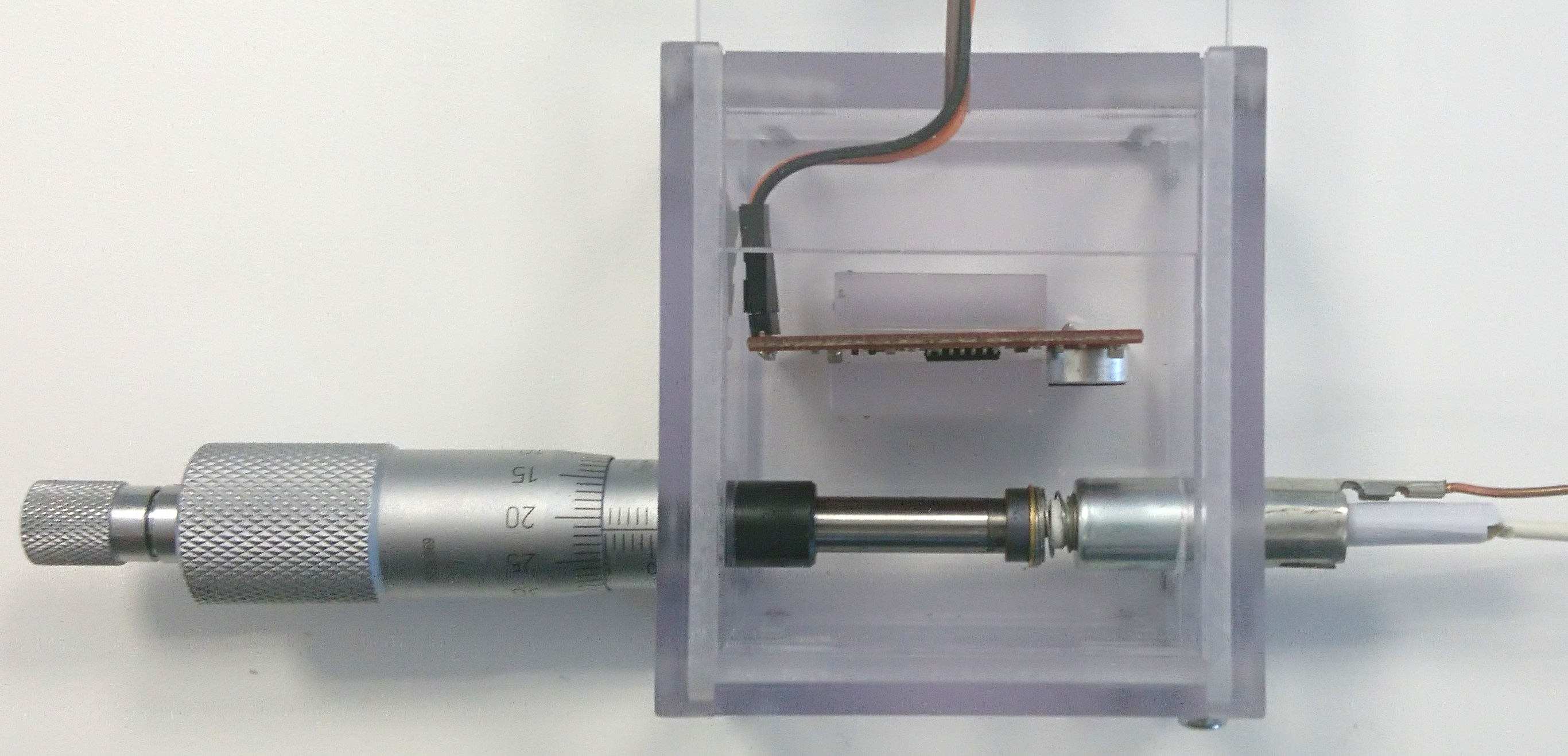}
	\caption{Linear-travel short-stroke solenoid valve (left) and experimental setup with the valve, a micrometer to limit the maximum gap and an electret microphone to measure the impact noise (right).}\label{valve}
\end{figure}

The device is a simple solenoid valve (Fig. \ref{valve}), which consists in a cylindrically symmetrical steel core, divided into a fixed component and a movable armature. The gap between the armature and the fixed core is constrained between a minimum and a maximum value. It has a single coil and a spring, which generate forces in opposing directions: in the making operation, the gap is closed via a magnetic force; whereas in the breaking operation, the gap is opened by reducing the magnetic force and allowing the spring to move the armature. The state-space functions and constants from \eqref{state_space_generic} are particularized for this device as follows,
\begin{subequations}
	\begin{align}
		f_v(\bm x)      & = \ddfrac{1}{m}\left( k_\mathrm{s}(z_\mathrm{s}-z) - c_\mathrm f \, v -\ddfrac{1}{2}\,\dfrac{\partial \mathcal R_\mathrm g(z)}{\partial z} \, {\alpha}^2 \right), \label{eq: f3 valve} \\
		G_v             & = 0,                                                                                                                                                                                   \\
		f_\alpha(\bm x) & = -\left(R \,\big( \mathcal R_\mathrm{c}(\alpha) + \mathcal R_\mathrm{g}(z) \big) \, \alpha \right)/(N^2 + R \, k_\mathrm{ec}), \label{eq: f_alpha}                                    \\
		G_\alpha        & = N/(N^2 + R \, k_\mathrm{ec}), \label{eq: G_alpha}
	\end{align}
\end{subequations}
where $\alpha$ is the magnetic flux, and the parameters are described and specified in Table \ref{table:parameters}.

Although not required for the optimal control problem, a complete model must also consider the position constrains between $z_\mathrm{min}$ and $z_\mathrm{max}$. This is accomplished by defining a hybrid automaton, which is described in \cite{Moya-Lasheras2017}, along with the magnetic reluctance expressions. Furthermore, the eddy current phenomenon is taken into account via the addition of the constant $k_\mathrm{ec}$ \cite{Ramirez-Laboreo2019b}.

To justify the use of the current as the control input (see Subsection \ref{di}), notice that the dynamic equation of the magnetic flux $\alpha$ with voltage as input (\ref{eq: f_alpha}) depends on the internal resistance $R$, which in turn depends greatly on temperature. Typically, the resistance dependence on temperature is negligible during an operation, but not after a large number of operation cycles. This makes the control with the voltage as input non-robust, i.e. a supplied voltage signal that achieves the desired behavior at a certain temperature is not guaranteed to work when that temperature changes. On the other hand, the relation between the current and the state of the device is not dependent on the resistance,
\begin{equation}\label{eq: output}
	{i = h(z, \alpha, \dot \alpha) = \big( \mathcal R_\mathrm{c}(\alpha) + \mathcal R_\mathrm{g}(z) \big) \, \alpha / N + k_\mathrm{ec} \, \dot \alpha.}
\end{equation}

Therefore, we propose to control the actuator by applying an optimal current signal. It is important to remark that the optimal control problem is still solved with the voltage as the control signal. The optimal current signal can then be easily calculated. Notice that the current derivative depends on $\ddot \alpha$ so, in order to accurately calculate it, an auxiliary variable $\alpha_2 = \dot \alpha$ should be added to the state vector $\bm x$. Alternatively, it can be approximated by setting $k_\mathrm{ec} = 0$. As the effect of the eddy currents is neglected, the solution is suboptimal. The error of the approximation will be illustrated in the following section. Note that this only affects this cost, for the rest of the problem $k_\mathrm{ec}$ is nonzero.

\begin{table}
	\centering
	\caption{{Solenoid valve parameters}} \label{table:parameters}
	\renewcommand{\arraystretch}{1.25} %
	\begin{tabular}{c c c c}
		\hline
		Parameter          & Description                  & Value                   & Units                    \\
		\hline \vspace{-0.35cm}                                                                                \\
		$ R $              & Resistance                   & $ 50 $                  & $ \mathrm{\Omega} $      \\
		$ N $              & Coil turns                   & $ 1.20 \times 10^{3}$   & $ \mathrm{1} $           \\
		$ m $              & Movable mass                 & $ 1.63 \times 10^{-3} $ & $ \mathrm{kg} $          \\
		$ k_\mathrm s $    & Spring stiffness coefficient & $ 6.18 \times 10^1 $    & $ \mathrm{N/m} $         \\
		$ z_\mathrm s $    & Spring equilibrium position  & $ 1.92 \times 10^{-2} $ & $ \mathrm{m} $           \\
		$ c_\mathrm f $    & Damping friction coefficient & $ 8.06 \times 10^{-1}$  & $ \mathrm{Ns/m} $        \\
		$ k_1 $            & Reluctance constant          & $ 4.41 \times 10^6 $    & $ \mathrm{H^{-1}} $      \\
		$ k_2 $            & Reluctance constant          & $ 3.80 \times 10^4 $    & $ \mathrm{Wb^{-1}} $     \\
		$ k_\mathrm{ec} $  & Eddy currents constant       & $ 1.63 \times 10^3 $    & $ \mathrm{\Omega^{-1}} $ \\
		$ z_\mathrm{min} $ & Minimum position             & $ 3.99 \times 10^{-4} $ & $ \mathrm{m} $           \\
		$ z_\mathrm{max} $ & Maximum position             & $ 1.60 \times 10^{-3} $ & $ \mathrm{m} $           \\
		\hline
	\end{tabular}
\end{table}

\subsection{Model identification}
The parameters are fitted with the following procedure: Firstly, the actuator is supplied with 26 square-wave voltage signals. Each one consists in $15 \ \mathrm{ms}$ of a positive voltage ($25, \, 26, \ldots, \ 50 \ \mathrm V$) for the making operation, followed by $15 \ \mathrm{ms}$ of $0 \ \mathrm V$ for the breaking operation. The measured voltage and current signals, $u_\mathrm{exp}$ and $i_\mathrm{exp}$, are sampled at $1 \ \mathrm{MHz}$. Secondly, the resistance $R$ and flux linkage $\lambda_\mathrm{exp}$ are estimated from  $u_\mathrm{exp}$ and $i_\mathrm{exp}$ according to the method explained in {\cite{Moya-Lasheras2017}}. Thirdly, the parameters are optimized by minimizing
\begin{equation}
	\hspace{-1mm}{\sum_{j=1}^{26} \left(\frac{(t_{\mathrm c, \mathrm{sim}, j} - t_{\mathrm c, \mathrm{exp}, j})^2}{{t_\mathrm{scale}}^2} + \frac{\sum_k (i_{\mathrm{sim},jk} - i_{\mathrm{exp},jk})^2}{\sum_k {i_{\mathrm{exp},jk}}^2} \right),}
\end{equation}
where the $t_{\mathrm c, \mathrm{sim}, j}$ and $t_{\mathrm c, \mathrm{exp}, j}$ are the experimental and simulated contact instants for the $j$th making operation, $i_{\mathrm{sim},jk}$ and $i_{\mathrm{exp},jk}$ are the $k$th sample of the $j$th cycle of the experimental and simulated current signals, and $t_\mathrm{scale} = 3 \ \mathrm{ms}$ is a scaling constant of the first addend.

The most straightforward way to identify the mechanical subsystem is to minimize the position errors {\cite{DiBernardo2012}}. Unfortunately, there is no sensor to measure the position. Nonetheless, if the voltage is constant, the contact instant $t_{\mathrm c, \mathrm{exp}, j}$ of each making operation is easily obtained from the current signal. It corresponds to the instant in which the current derivative changes abruptly from negative to positive. The breaking contact instants cannot be precisely derived, so they are not used. On the other hand, the simulated contact instants $t_{\mathrm c, \mathrm{sim}, j}$ are obtained directly from the simulated position $z_\mathrm{sim}$ and velocity $v_\mathrm{sim}$ which, for a set of parameters, are calculated by solving the differential equations \eqref{x1x2}, \eqref{eq: dot v}, and utilizing as input of this subsystem $\alpha = \lambda_\mathrm{exp} / N$. In those simulated or experimental cases where the generated magnetic force is insufficient to move the armature, their contact instants are set to zero.\looseness=-1

For the identification of  the electromagnetic subsystem, the simulated magnetic flux is obtained by solving \eqref{eq: dot alpha}, using \mbox{$z = z_\mathrm{sim}$}, $v = v_\mathrm{sim}$ and $u = u_\mathrm{exp}$ as input signals. Then the current $i_\mathrm{sim}$ is calculated according to \eqref{eq: output}.

\begin{figure}[t]
	\includegraphics{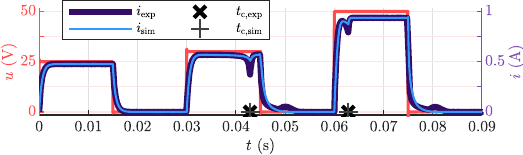}
	\caption{Experimental and simulated results for voltage pulses of $25$, $30$ and $50 \ \mathrm V$. In the first one, the valve does not move.}\label{fig: validate}
\end{figure}

Fig. \ref{fig: validate} depicts the measurements for three different voltage pulses, used for identification. For comparison purposes, the simulated results after the identification are also displayed. In the first pulse, the $25 \ \mathrm{V}$ is not enough to move the valve. In the following ones the contact instants, both experimental and simulated, are shown. Notice the great difference between both: one contact instant is delayed $13 \ \mathrm{ms}$ from the start of the voltage pulse, and the other only $3 \ \mathrm{ms}$.

\section{Analysis}
In this section, simulated and experimental tests are performed to compare our proposal, the probability-based optimal solution, with a state-of-the-art optimal solution in which the position randomness is not taken into account. The trajectories are obtained by solving the BVP with the MATLAB function \texttt{bvp4c} \cite{Kierzenka2001}, integrating with an adaptive step size.

\subsection{Optimal voltage signal}\label{Problem formulation and solver} \label{formulation}
According to \eqref{eq: minH}, the optimal control $u^*$ for this particular case is defined as
\begin{equation}\label{eq: minH_v2}
	u^*(\bm x^*, \bm \lambda^*)  = \argmin_{u_-\leq u \leq u_+} \mathcal H(\bm x^*, \bm \lambda^* , u),
\end{equation}
where $u_-$ and $u_+$ are the lower and upper limits of the optimal voltage input. In the case the voltage is used as the input in the implementation, they could be set directly to $u_\mathrm{min}$ and $u_\mathrm{max}$ respectively, which would be the minimum and maximum supply voltage. However, if the current is used as input, the real voltage changes with the resistance, as noted in Subsection {\ref{subsec: model}}. This means that the values of  $u_-$ and $u_+$ must be selected conservatively to guarantee that the designed current is actually obtainable with a voltage between $u_\mathrm{min}$ and $u_\mathrm{max}$ and a real resistance $R_\mathrm{real}$ ranging from $R_\mathrm{min}$ to $R_\mathrm{max}$. Given the electrical circuit equation,
\begin{equation}
	{u = R \, i + N \, \alpha,}
\end{equation}
it is possible to determine the worst-case scenarios,
\begin{equation}
	{- R_\mathrm{real} \, i + u_\mathrm{min} \leq -R \, i + u^* \leq - R_\mathrm{real} \, i + u_\mathrm{max},}
\end{equation}
being $R$ the resistance used in the optimal control equations. Then, the conditions for bounding the optimal input are
\begin{align}
	{u_+} & {\leq u_\mathrm{max} + \min\big((R - R_\mathrm{real}) \, i\big),} \\
	{u_-} & {\geq u_\mathrm{min} + \max\big((R - R_\mathrm{real}) \, i\big),}
\end{align}

Provided that the current is bounded such that $i \in [0, \ i_\mathrm{max}]$, the maximum value can be defined conservatively as $i_\mathrm{max} = u_+/R$. In that case,
\begin{align}
	{u_+} & {\leq u_\mathrm{max} \, R / R_\mathrm{max},}          \\
	{u_-} & {\geq u_\mathrm{min} + u_+ (R - R_\mathrm{min}) / R.}
\end{align}

To solve \eqref{eq: minH_v2} algebraically, an auxiliary variable is defined,
\begin{align}
	q^*(\bm x^*, \bm \lambda^*) & = \argmin_u \mathcal H(\bm x^*, \bm \lambda^* , u) \nonumber                                  \\
	                            & = u \;\; \mathrm{s.t.} \;\; \partial \mathcal H(\bm x^*, \bm \lambda^* , u) / \partial u = 0,
\end{align}
which is unique because $\partial \mathcal H / \partial u$ is linear in $u$. Then, $u^*$ can be obtained by simply saturating $q^*$ between the control limits,
\begin{equation}
	\hspace{-1mm}
	u^*(\bm x^*, \bm \lambda^*) = \left\lbrace \hspace{-1mm}
	\begin{array}{rr}
		q^*(\bm x^*, \bm \lambda^*), & \hspace{-1mm} u_-\leq q^*(\bm x^*, \bm \lambda^*) \leq u_+, \\
		u_-,                         & \hspace{-1mm} q^*(\bm x^*, \bm \lambda^*)<u_-,              \\
		u_+,                         & \hspace{-1mm} q^*(\bm x^*, \bm \lambda^*)>u_+.              \\
	\end{array}\right.
\end{equation}

To prove it, notice that $q^*$ is the global minimum of $\mathcal H$ and $\partial^2 \mathcal H / \partial u^2$ does not depend on $u$. Therefore, $\partial \mathcal H / \partial u > 0$ for any $u<q^*$ and $\partial \mathcal H / \partial u < 0$ for any $u>q^*$. Thus, for any $u \in (u_-, u_+)$ and $q^* \not\in [u_-, u_+]$,
\begin{subequations}
	\begin{align}
		q^* \! > \! u_+ \! > \! u \, \Rightarrow \, \mathcal H(\cdot,\cdot,u_+) \! < \! \mathcal H(\cdot,\cdot,u) \, \Rightarrow \, u^*=u_+, \\
		q^* \! < \! u_- \! < \! u \, \Rightarrow \, \mathcal H(\cdot,\cdot,u_-) \! < \! \mathcal H(\cdot,\cdot,u) \, \Rightarrow \, u^*=u_-.
	\end{align}
\end{subequations}

Analogously to the acceleration in Subsection {\ref{expected acceleration}}, the saturation of $q^*$ should be approximated with a differentiable function.

\subsection{Optimization specifications}
The parameters that specify the optimal control solution are summarized in Table \ref{table:param_op}. They correspond to a making operation, in which {$z(t_\mathrm 0) > z(t_\mathrm f)$}. To account for the uncertainty, the contact position $Z_\mathrm{c}$ is considered a normal deviate, with a mean $\mu_z$ and a variance ${\sigma_z}^2$. To analyze the probability-based optimal solutions (POS), they are compared with an energy-optimal solution (EOS) for soft landing with no uncertainty considerations. For the latter case, there are essentially two differences: in the boundary conditions for $t_\mathrm f$, which forces the velocity and acceleration to be zero,
\begin{align}
	f_v(\bm x^*(t_\mathrm f)) & = 0, & v^*(t_\mathrm f) & = 0;
\end{align}
and in the cost functional, which corresponds to an energy-optimal control problem,
\begin{align}
	{J = \int_0^{t_\mathrm f} u^2(t) \, \mathrm dt.}
\end{align}

\begin{table}
	\centering
	\caption{{Optimization parameters}} \label{table:param_op}
	\renewcommand{\arraystretch}{1.25} %
	\begin{tabular}{c c c c}
		\hline
		Parameter                   & Description               & Value                         & Units            \\
		\hline \vspace{-0.35cm}                                                                                    \\
		$ [u_-, \, u_+]$            & Voltage bounds            & $ [-45, \, 45] $              & $ \mathrm{V} $   \\
		$ z_\mathrm{0} $            & Initial position          & $ 1.60 \times 10^{-3} $       & $ \mathrm{m} $   \\
		$ z_\mathrm{f} $            & Final position            & $ 3.99 \times 10^{-4} $       & $ \mathrm{m} $   \\
		$ t_\mathrm f $             & Final time                & $ 3.5 \times 10^{-3} $        & $ \mathrm s $    \\
		$ \mu_z $                   & Expected contact position & $ 3.99 \times 10^{-4} $       & $ \mathrm{m} $   \\
		$ {\sigma_z}^2 $            & Contact position variance & $ 4 \times 10^{-10} $         & $ \mathrm{m^2} $ \\
		$ \{w_1, \, w_2, \, w_3 \}$ & Cost weights              & $ \{ 10^6, \, 10^3, \, 1 \} $ & $ \mathrm{1} $   \\
		\hline
	\end{tabular}
\end{table}

To make a fair comparison between both cases, we set $z_\mathrm f = \mu_\mathrm z$. This means there is a $50 \ \%$ probability of no contact in the simulated time interval, i.e.,
\begin{equation}
	P(0 \leq T_\mathrm c \leq t_\mathrm f) = 0.50.
\end{equation}

Furthermore, the bounced velocity---needed only for the calculation of the expected bounced acceleration---is chosen conservatively. From {\eqref{eq: f3 valve}}, it is easy to see that the acceleration increases if the velocity decreases, so the worst-case scenario corresponds to $v_\mathrm b = 0$.

\subsection{Comparison via simulation}
\begin{figure}[h!t]
	\centering\includegraphics{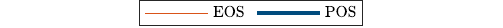}\vspace{-2mm}
	\\
	\subfloat[Voltage.]{\includegraphics{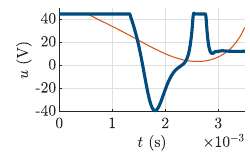}\label{fig:voltage}}
	\subfloat[Current.]{\includegraphics{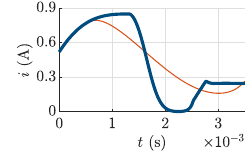}\label{fig:current}}
	\\
	\subfloat[{Current derivative.}]{\includegraphics{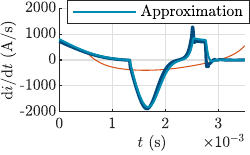}\label{fig:dcurrent}}
	\subfloat[{Magnetic force.}]{\includegraphics{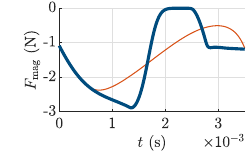}\label{fig:fmag}}
	\\
	\subfloat[Magnetic flux.]{\includegraphics{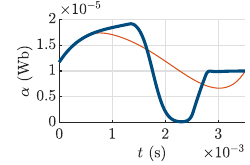}\label{fig:flux}}
	\subfloat[Position.]{\includegraphics{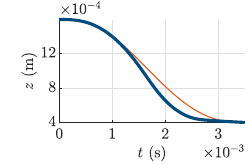}\label{fig:position}}
	\\
	\subfloat[Contact velocity as a function of the contact position. EOS: {${\mathrm{E}[V_\mathrm{c}] = -0.0998 \ \mathrm{m/s}}$}. POS: {${\mathrm{E}[V_\mathrm{c}] = -0.0468 \ \mathrm{m/s}}$}.]{\includegraphics{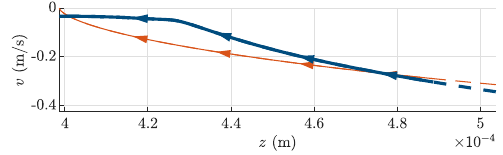}\label{fig:z-v}}
	\\
	\subfloat[Contact acceleration as a function of the contact position. EOS: {${\mathrm{E}[A_\mathrm{c}] = 357.40 \ \mathrm{m/s}^2}$}. POS: {${\mathrm{E}[A_\mathrm{c}] = 76.52 \ \mathrm{m/s}^2}$}.]{\includegraphics{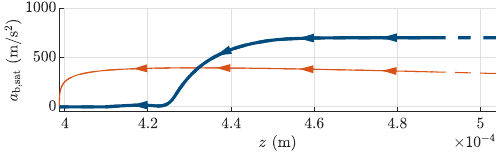}\label{fig:z-a}}
	\\
	\subfloat[Contact position probability density function.]{\includegraphics{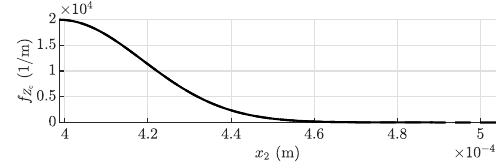}\label{fig:f_Z}}
	\caption{Comparison of simulated results from EOS and POS.}\label{fig:compare}
\end{figure}

The simulated results of our proposal and state-of-the-art optimal solutions are presented in Fig. \ref{fig:compare}. Although our trajectory-optimal proposal presents a term for the minimization of the current derivative, its weight is purposely set to be much smaller than the others, in order to prioritize the minimization of the expected contact velocity and acceleration. For that reason, the current (Fig. \ref{fig:current}) signal, as well as the voltage (Fig. \ref{fig:voltage}), are steeper than the ones from EOS. Note also that both voltage signals are saturated to $u_+ = 45 \ \mathrm{V}$.

Note that the model takes into account the eddy currents, but it is neglected in the calculation of $V_3$ \eqref{V_3}. In Fig. \ref{fig:dcurrent}, the approximated current derivative---where $k_\mathrm{ec} = 0$---is also displayed. There is a noticeable, albeit small, error.

As seen in Fig. \ref{fig:position}, the position of the EOS has a steadier transition than POS, which shifts abruptly toward the final position, but slows down quickly when the probability of contact stops being negligible. The expectations of the velocity and acceleration in the case of contact are therefore smaller. This improvement comes at the expense of an energy consumption increase. The EOS consumption is $47.0 \ \mathrm{mJ}$, better than POS, $52.8 \ \mathrm{mJ}$, which is 12 \% larger.

To better comprehend the advantage of minimizing the expected contact velocity and acceleration, it is useful to visualize the velocity $v$ and acceleration $a_\mathrm{b,sat}$ trajectories with respect to position $z$, as in the state planes presented in Figs. \ref{fig:z-v} and \ref{fig:z-a}. The arrows show the direction of their evolution over time, as $z$ approaches $z_\mathrm f$. Notice that, instead of showing the complete trajectory, the horizontal $z$-axis is limited to $5 \times 10^{-4} \ \mathrm m$ because larger positions have a negligible probability of contact. Note also that acceleration $a_\mathrm{b,sat}$ cannot be negative, as explained in Subsection \ref{expected acceleration}. These graphics represent the contact velocity and acceleration for every possible contact position. The probability density function of the contact position is also presented in Fig. \ref{fig:f_Z}. The EOS velocity and acceleration are exactly zero in the expected contact position, $z = \mu_z = 3.99 \times 10^{-4} \ \mathrm m$, but their values change steeply as the position does. POS, instead, keeps a small and steady velocity and acceleration in the position interval in which the probability of contact is significant. This behavior results in considerably better expected contact velocities and accelerations.\looseness=-1

\begin{figure}[t]
	\centering\includegraphics{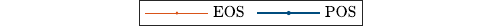}\vspace{-2mm}

	\subfloat[\label{fig:v make}]	{\includegraphics{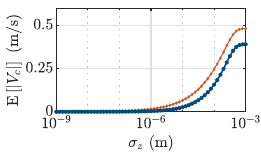}}
	\subfloat[\label{fig:v break}]	{\includegraphics{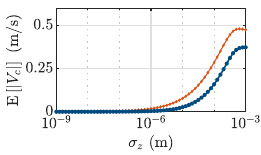}}

	\subfloat[\label{fig:a make}]	{\includegraphics{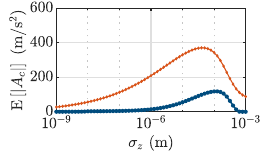}}
	\subfloat[\label{fig:a break}]	{\includegraphics{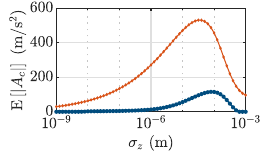}}
	\caption{Absolute values of expected contact velocity (top) and acceleration (bottom) in making (left) and breaking (right) operations, for different {$\sigma_z$}.}\label{compare_SD}
\end{figure}

Additionally, to make a more complete comparison, multiple simulations are performed by modifying the contact position variance, for both operations, while the rest of the parameters are kept as stated in the previous subsection. As the range of standard deviations utilized in the simulations is very wide, the horizontal axis is presented in logarithmic scale. The results, displayed in Fig. \ref{compare_SD}, show a persistent improvement of POS in the expected velocities and accelerations with respect to EOS, for both types of operations. Unsurprisingly, as the uncertainty decreases, i.e. $\sigma_z$ is reduced, the expectations of velocity and acceleration in the contact tend to zero for both solutions. This is more prominent in the case of the expected contact velocities, which are very close to zero for both methods if $\sigma_z<10^{-7} \ \mathrm{m}$ (see Figs. \ref{fig:v make} and \ref{fig:v break}). However, the expected accelerations from the EOS solutions are still substantial for small values of $\sigma_z$, whereas the ones from the proposed solutions are insignificant (see Figs. \ref{fig:a make} and \ref{fig:a break}).

\subsection{Comparison via experimentation}

To validate the improvement in a real application, the optimal solutions are applied to the presented solenoid valve. The lack of a position sensor is an important limitation for the experimental testing, but instead it is possible to measure the impact noises with an electret microphone (Fig. \ref{valve}).

Three different current signals for the making operation are alternately applied to the valve, 500 times each. The first one is simply set to $0.8 \ \mathrm A$ (no control), and serves as a reference for the other two. The second and third ones are the generated EOS and POS current signals, which were also used in the simulations (see Fig. \ref{fig:current}). For each one, there is a constant-slope transition from $0 \ \mathrm A$ to the initial current value in $ 2 \ \mathrm{ms}$ and another constant-slope transition from the final current value to $0.8 \ \mathrm A$ in $ 2 \ \mathrm{ms}$. The current is then kept at $0.8 \ \mathrm A$ for a sufficiently long period of time to ensure the commutation and to completely measure the audio with the microphone. We focus only on the making operations, which present the most notable impact noises in this device.

To process the voltage signals from the microphone, the following energy is obtained for each one,
\begin{equation}
	E_\mathrm s = \int_{t_0}^{t_0 + \Delta t} {u_\mathrm{audio}}^2(t) \, \mathrm dt,
\end{equation}
where $t_0$ is established as the first instant $t$ where $ u_\mathrm{audio}(t) > \mathrm{max}( u_\mathrm{audio})/5$ and $\Delta t = 0.01 \ \mathrm s$. The energies are then normalized by dividing each one by $1.52 \times 10^{-3} \ \mathrm V^2 \mathrm s$, which is the average of the 500 ones with no control (its relative standard deviation is $0.1561$).

\begin{figure}
	\includegraphics{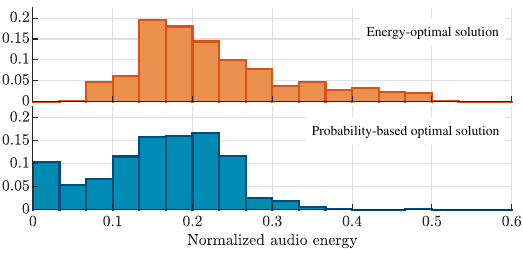}
	\caption{Relative frequency histograms of the normalized energies from the audio signals. Comparison between EOS and POS.}\label{fig: hist}
\end{figure}

The results from the optimal control solutions are condensed in the histograms shown in Fig. {\ref{fig: hist}}. Both reduce considerably the impact sound with respect to the average operation with no control. The results from POS are appreciably better, with an average of $0.161$ and a standard deviation of $ 0.0795 $, in contrast with the average of $0.2242$ and standard deviation of $0.0947$ from the energy-optimal solution.

\section{Conclusions}\label{Conclusions}
In this paper, we have proposed a new optimal approach to design soft-landing trajectories of actuators, specifically simple short-stroke devices that are difficult to accurately control. For the case of reluctance actuators, the advantage of using the current as input was discussed and a term was added to the cost functional to minimize the square of the current derivative. It is also possible to include additional terms to the cost functional if it is required to optimize other concepts, e.g. the contact time or the power consumption.

Although the contact position is considered a random variable, the system dynamics is still defined as a deterministic model, which permits formulating and solving a regular optimal control problem. In practice, the random contact position deviation can be magnified to compensate for uncertainty of the model. The experimental results show the improvement of considering uncertainty in the contact position, even though there are other sources of uncertainty that are not directly taken into account. Also, they help to highlight how challenging these types of devices are to soft-landing control when there is no position sensor or observer. Notice that, even though the same current signal is applied to the device repeatedly, the resulting impact noise has a great dispersion.

\section*{Acknowledgment}
The authors want to thank S. Llorente for his support and BSH Home Appliances Spain for supplying the solenoid valve and other equipment used in this research.

\bibliographystyle{IEEEtran}

\end{document}